\documentclass[twocolumn,showpacs,preprintnumbers,amsmath,amssymb,floatfix]{revtex4}
\input{epsf}

\usepackage{graphicx}% Include figure files
\usepackage{dcolumn}% Align table columns on decimal point
\usepackage{bm}% bold math

\begin{document}

\preprint{APS/123-QED}

\title{Entanglement between atomic condensates in an optical lattice: effects of interaction range }
\author{H. T. Ng and K. Burnett}
\affiliation{{Clarendon Laboratory, Department of Physics,
University of Oxford, Parks Road, Oxford OX1 3PU, United Kingdom}}
\date{\today}

\begin{abstract}
We study the area-dependent entropy and two-site entanglement for
two state Bose-Einstein condensates in a 2D optical lattice. We
consider the case where the array of two component condensates
behave like an ensemble of spin-half particles with the interaction
to its nearest neighbors and next nearest neighbors. We show how the
Hamiltonian of their Bose-Einstein condensate lattice with
nearest-neighbor and next-nearest-neighbor interactions can be
mapped into a harmonic lattice.  We use this to determine the
entropy and entanglement content of the lattice.
\end{abstract}


\maketitle
\section{Introduction}
Since Bose-Einstein condensates (BEC's) of alkali gases were first
observed in a magnetic trap \cite{Bradley}, the experimental study
of BEC's has grown rapidly. In particular, multi-component BEC's
have been realized using the different hyperfine states of
${}^{87}$Rb \cite{Myatt} and spinor condensates using different
Zeeman states of sodium $F=1$ \cite{Stenger}. In addition,
interspecies interactions give rise to interesting phenomena such as
phase-separation \cite{Ho}. Besides this, the quantum phase
transition (QPT) \cite{Sachdev} from a superfluid phase to a Mott
insulator phase has been observed using atoms  in optical lattices
\cite{Greiner}.   BEC's in optical lattices are also being used to
study aspects of quantum entanglement \cite{Nielsen} in many-body
systems because of their controllability \cite{Mandel}. Indeed,
evidence of multi-particle entanglement has been seen in atoms with
two internal states prepared in Mott insulator in optical lattices
\cite{Mandel}.

In this paper, we study the entanglement in the spin waves of a
two-component BEC \cite{Myatt} in a 2D optical lattice
\cite{Auerbach} in which the condensates at each site behave like a
``spin magnet'' and interact with each other to its nearest
neighbors and next nearest neighbors. The low lying excitation of
this system are spin waves. Most importantly, a ``long-ranged''
interaction can in principle be produced by dipole-dipole
interactions. With this in mind we note that Bose-Einstein
condensation of chromium, with a high magnetic dipole-dipole
interaction, has recently been realized \cite{Griesmaier}. This
dipole-dipole interaction strength should also be tunable by
magnetic fields, or by engineering the geometry of the trap
\cite{Giovanazzi}. Such controllable interspecies, short- and
long-ranged interactions are important in producing the
multi-particle entangled states.

The study of the entanglement measures leads to a novel perspective
on the structure of ground state and its quantum critical behavior.
We shall show that this ``BEC lattice'' is equivalent to a set of
harmonic oscillators in the low-excitation regime. This means that
the well-known analysis of a quantum harmonic lattice can be used to
investigate entanglement in this many-body system. We shall further
see that the entropic measure of entanglement is useful to
understand our system's properties of area dependence which used in
some parts of quantum field theory \cite{Bombelli}. In the case of
the Klein-Gordon (KG) field, the entropy of the field is obtained by
tracing over the variables outside a region under consideration and
the entropy found to be directly proportional to the boundary area
of inside region. This was shown by Bombelli {\it et al.} and
Srednicki by considering a free, massless and scalar KG field, and
which is also equivalent to the ground state of a coupled of
harmonic oscillators \cite{Bombelli}.

An analysis of the entropy and entanglement in a 1D harmonic lattice
system was from the viewpoint of quantum information theory was
given in the reference \cite{Audenaert}, along with the
generalization to the 2D and 3D cases. The area-dependence of the
entropy \cite{Plenio} and the quantum correlations \cite{Schuch} in
a harmonic lattice including some aspects of the critical behavior
were also studied.  Nevertheless, the general relationship between
area-dependent entropy and the nature of QPT is not yet fully
understood \cite{Plenio}.  We will examine an aspect of this
relationship and provide some new insights.

Two-site entanglement is also a most useful quantity with which to
examine the nature of quantum correlations in a lattice system.  For
the case of an infinite spin-chain, with nearest neighbor
interaction, the critical behavior of two-site entanglement was
given in references \cite{Osterloh}.  It is clear from these studies
that two-site entanglement shows the non-local nature of a many-body
system close to QPT.

This paper is organized as the follows: In Sec. II, we introduce the
system of two-component dipolar condensates trapped in a
two-dimensional optical lattice.  In Sec. III, we make the Gaussian
approximation to the system and show that the system can be
represented in terms of harmonic oscillators.  The quantum phases
(QPT) of the system are then determined. In Sec. IV, we study the
area-dependent entropy and its behavior closing to a QPT.  In Sec.
V, we investigate two-site entanglement and then we give out a
conclusion.

\begin{figure}[ht]
\includegraphics[height=3cm]{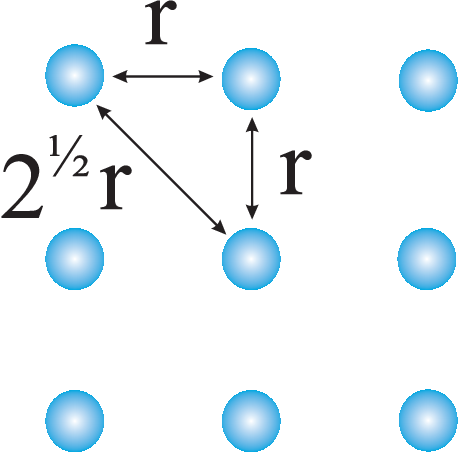}
\caption{ \label{fig1} The configuration of the 2D lattice with
distance between the nearest neighbor $r$, and with the next-nearest
neighbor $2^{1/2}r$.}
\end{figure}

\section{System}
We consider a two-component BEC trapped in a square optical lattice
with $M{\times}M$ sites as shown in Fig. \ref{fig1}. The atoms
involved have two different magnetic states with tunable magnetic
dipolar interaction strength \cite{Griesmaier,Giovanazzi,Goral} in
which the dipolar interaction potential energy is
\begin{eqnarray}
U(r)=-\frac{\mu_0}{4\pi}\frac{3(\mathbf{m_1}\cdot\hat{r})(\mathbf{m_2}\cdot\hat{r})-(\mathbf{m_1}{\cdot}\mathbf{m_2})}{r^3},
\end{eqnarray}
for $\mathbf{m_1}$ and $\mathbf{m_2}$ are the two magnetic dipole
moments at $\mathbf{r_1}$ and $\mathbf{r_2}$ respectively, and
${r}=|\mathbf{r_1}-\mathbf{r_2}|$ is the distance between two dipole
moments and $\hat{r}=(\mathbf{r_1}-\mathbf{r_2})/r$ is an unit
vector.

Moreover,  we assume that these two internal states of atoms can be
coupled by applying an external field. We adopt the single-mode
approximation \cite{Milburn} to the condensates trapped deeply in
each site in which the condensates can be described by the localized
mode functions associated with the potential wells. The Hamiltonian
of the system is then given by ($\hbar=1$),
\begin{eqnarray}
H&=&H_{\rm sys}+H_{\rm ext},
\end{eqnarray}
\begin{widetext}
\begin{eqnarray}
H_{\rm sys}&=&
\sum_{<i,j>}g^{ij}_aa^\dag_ia_ia^\dag_ja_j+g^{ij}_bb^\dag_ib_ib^\dag_jb_j+g^{ij}_{ab}(a^\dag_ia_ib^\dag_{j}b_{j}+a^\dag_ja_jb^\dag_{i}b_{i})
+\Omega^{ij}_a(a^\dag_ia_j+a^\dag_ja_i)
+\Omega^{ij}_b(b^\dag_ib_j+b^\dag_jb_i),\nonumber\\
&&\sum_{i}\kappa^i_a(a^\dag_ia_i)^2+\kappa^i_b(b^\dag_ib_i)^2
+\kappa^i_{ab}(a^\dag_ia_ib^\dag_ib_i)\\
H_{\rm ext}&=&\sum_i\omega^i(a^\dag_ib_i+b^\dag_ia_i)/2,
\end{eqnarray}
\end{widetext}
where $a_i$ and $b_i$ are the annihilation operators of components A
and B in the $i$-th site. The interaction parameters
$\kappa^i_a(\kappa^i_b)$, $\kappa^i_{ab}$ and $\omega_i$ are
respectively the intra-component interaction, the inter-component
interaction and the coupling rate between the two internal states in
the $i$-th site. The hopping and the dipole-dipole interaction
strengths between the $i$-th site and the $j$-th site are denoted by
$\Omega^{ij}_a(\Omega^{ij}_b)$ and $g^{ij}_a$($g^{ij}_b$) for the
component A(B) respectively, and $g^{ij}_{ab}$ is the dipolar
strength between the two different component condensates. We
consider the atoms in each site to interact with the nearest
neighbors and the next-nearest neighbors only as shown in Fig.
\ref{fig1}.  We neglect the interaction between atoms with the
larger separations for simplicity and intend to return to their role
in the future work.  We should emphasize, however, that we do find
novel physics associated with the combination of on-site plus
neighbor-neighbor interaction.

For the purposes of discussion, we consider the interaction
parameters and the number of atoms to be the same for each site,
i.e., $\omega^i=\omega$,
$\kappa=(\kappa^i_{a}+\kappa^i_{b}-\kappa^i_{ab})/4$ and $N^i=N$.
Moreover, the atoms between the nearest neighbor with the same
dipolar strength is considered, for $g^{ij}_a=g_a$, $g^{ij}_b=g_b$,
$g^{ij}_{ab}=g_{ab}$ and $i,j$ are two indices for two nearest
neighbors. In the Mott-insulator limit, $\Omega{\ll}{\kappa}$, the
tunneling between the sites is negligible and hence the trapped
two-component condensates in each site can be regarded as an
ensemble of pseudo spin-half particles. For this case, we can write
the Hamiltonian in terms of angular momentum operators in the
following form (omit the constant):
\begin{eqnarray}
\label{Hamiltonian1} H_1&=&\sum_i\omega{J^i_z}+4{\kappa}J^{i2}_x
+\sum_{<ij>}g^{ij}J^i_xJ^j_x,
\end{eqnarray}
Here $J^i_x=(a^\dag_ia_i-b^\dag_ib_i)/2$,
$J^i_y=(a^\dag_ib_i-a_ib^\dag_i)/2i$,
$J^i_z=(a^\dag_ib_i+a_ib^\dag_i)/2$, $N=a^\dag_ia_i+b^\dag_ib_i$ is
the total number of atoms in each site and the dipolar strength
$g^{ij}$ of the nearest neighbor and the next-nearest neighbor
interaction are $g$ and $2^{-3/2}g$ respectively, for
$g=g_a+g_b-2g_{ab}$.  The factor $2^{-3/2}$ appears in the next
nearest neighbor comes from the spatial dependence of the square
lattice. We also assume that $|\kappa^i_a-\kappa^i_b|N$ is very
small so that we have neglected the linear terms $J^i_x$. This shows
that atomic dipole-dipole interaction systems will be a good place
to study the combined effects of nonlinear on-site interaction with
interaction between neighbors.

\section{Quantum phases}
We assume a sufficiently large external field is applied to the
system such that $\omega\gg\kappa,g$ and in the order of
${\kappa}N$.  This means that the spin on each site is initially
aligned to the negative $z$-direction of the angular momentum basis.
The small rotation around this negative $z$-direction can be
described as the motion of a harmonic oscillator in the phase-space
plane.  To describe this harmonic motion, we can apply the
Holstein-Primakoff transformation (HPT) \cite{Holstein} to map the
spin operators into the harmonic oscillators, where
$J^i_+=c^\dag_i\sqrt{N-c^\dag_ic_i}$,
$J^i_-=\sqrt{N-c^\dag_ic_i}c_i$ and  $J^i_z=(c^\dag_ic_i-N/2)$. In
the low degree of excitation regime,
$\langle{c^\dag_ic_i}\rangle/N{\ll}1$, the Hamiltonian can be
approximated thus:
\begin{eqnarray}
\label{Hamiltonian2}
H_{2}&=&\sum_i\omega(c^\dag_ic_i)+\kappa{N}(c^\dag_i+c_i)^2\nonumber\\
&&+\frac{N}{4}\sum_{<i,j>}g^{ij}(c^\dag_i+c_i)(c^\dag_j+c_j).
\end{eqnarray}
This effective Hamiltonian $H_2$ is the zero order approximation of
this exact Hamiltonian $H_1$ and gives a description of spin waves
\cite{Auerbach}. In the thermodynamic limit, this approximation
becomes exactly equivalent to the system $H_1$. The exact numerical
solution of this multi-spin system even for just a few sites is
extremely difficult.  This analytical although approximate solution
is therefore extremely valuable and provides important insight into
the physics of this multi-spin system. Indeed, the approximation
turns out to be valid in the two-site case even if the number of
atoms $N$ is in the order of several hundred \cite{Ng}.  We
immediately see from this that the ground state and low-lying
eigenstates of our system will behave like  a set of coupled
harmonic oscillators.  We can see this by representing the operators
in the position-momentum space:
$q_i=(c^\dag_i+c_i)/{\sqrt{2\omega}}$ and
$p_i=i\sqrt{\omega}(c^\dag_i-c_i)/\sqrt{2}$. Then, the Hamiltonian
can be written explicitly in terms of a set of coupled harmonic
oscillators:
\begin{eqnarray}
H_2'&=&\frac{1}{2}\sum_ip^2_i+\frac{1}{2}\sum_{<i,j>}q_iV_{ij}q_j.
\end{eqnarray}
Here, $V_{ij}$ is the potential matrix that expresses the
interaction strength between the oscillators $q_i$ and $q_j$. In the
infinite lattice limit, the spectrum of the Hamiltonian can now be
found by using a two-dimensional Fourier transform.

The onset of a second-order QPT can now be determined by simply
finding when the energy gap $\Delta$, i.e., the energy difference
between the first excited and ground states of the system, vanishes
\cite{Sachdev}. This model is now directly solvable, and so the spin
wave modes and the energy gap $\Delta$ can be evaluated explicitly.
The critical coupling $g_c$ is $(\omega+4\kappa{N})/N(4-\sqrt{2})$.
As $g$ approaches to $g_c$, the lowest excitation energy gap is
given by $\Delta=[\omega{N}(4-\sqrt{2})]^{1/2}|g-g_c|^{1/2}$. If $g$
is greater than the critical value $g^c$, then the excitation gap
$\Delta$ becomes complex, the low-lying excitations approximation
breaks down, and the Gaussian approximation thus fails. This implies
the pseudo spin-half particles in each site will be ``polarized'' to
some extent which can be viewed as ``quantum magnetization'' of the
spin-half atoms \cite{Auerbach}. We intend to examine this issue in
our future work.

\begin{figure}[ht]
\includegraphics[height=6cm]{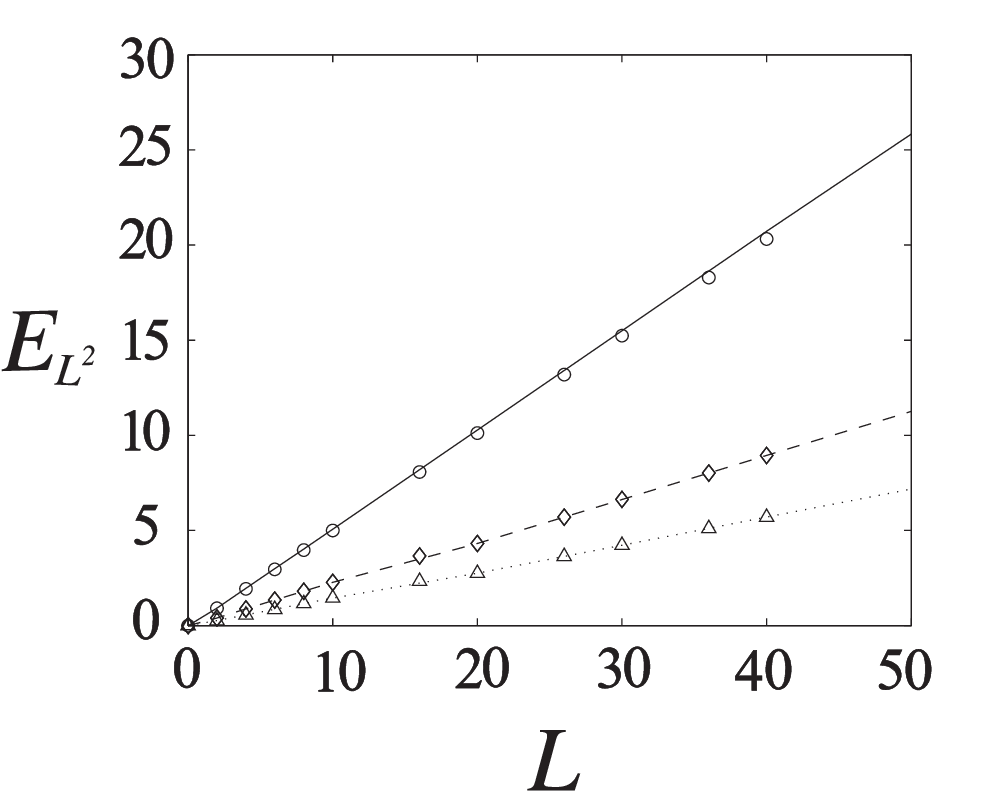}
\caption{\label{fig2} The  entropy $E_{L^2}$ as a function of ${L}$
is shown, for $\omega=500\kappa$, $N=1000$ and the lattice size
$M{\times}M$ are 6400.  For the infinite (finite) case, the
different coupling strengths $g/\kappa$ is approximate to
$g_c/\kappa=1.74028\ldots$ up to 11 decimal place, 1.5, 1.25 are
shown, where are denoted by solid line (empty circle), dashed line
(empty square) and dotted line (empty up-triangle) respectively. }
\end{figure}

\section{Area-dependent Entropy}
We can now examine the entanglement properties of the ground state
of our system.  As the ground state is a Gaussian state, its
$2M\times2M$ density matrix can be completely determined from the
second order moments
$\langle{X_iX_j+X_jX_i}\rangle-2\langle{X_i}\rangle\langle{X_j}\rangle$,
where $X_i$ are the quadrature variables $q_i(p_i)$ and
$i=1,\ldots,M$.  The density matrix of the system $\rho$ can be
expressed purely in terms of the covariances
$\langle{X_iX_j}\rangle$, since $\langle{X_i}\rangle=0$ in the
ground state.

We now investigate the area-dependent properties of the entropy of
the bipartite entanglement of two sets of harmonic oscillators. We
consider the $M\times{M}$ square harmonic lattice is bisected into a
$L\times{L}$ square lattice and the remaining $M^2-L^2$ lattice as
``inside'' and ``outside'' parts respectively, say 1 and 2.
Moreover, we consider a $L{\times}L$ ``inside'' square lattice which
is located at the center of the whole lattice whereas the rest of
oscillators are called the ``outside'' part. The bipartite
entanglement can be determined by the von-Neumann entropy
$E_{L^2}(\rho)$ of the reduced density matrix which is given by
\begin{eqnarray}
E_{L^2}(\rho)&=&-{\rm Tr}(\rho_1\log_2\rho_1),
\end{eqnarray}
where $\rho_1={\rm Tr}_2(\rho)$ is the reduced density matrix of
$\rho$ by tracing out the ``outside'' subsystem 2. The reduced
density matrix $\rho_1$ can be obtained by including the position
and momentum correlations in the set of modes 1 only. It is given by
\begin{equation}
\rho_1 = \left( {\begin{array}{clcr}
Q & 0 \\
0 & P
\end{array}} \right),
\end{equation}
where $Q$ and $P$ are the position and momentum covariance matrices
with the matrix elements $Q_{ij}=\langle{q_iq_j}\rangle$ and
$P_{ij}=\langle{p_ip_j}\rangle$ of the subsystem 1 respectively. The
entropy measure of this bipartite entanglement between two regions
$E_{L^2}$ is then found to be \cite{Plenio}
\begin{eqnarray}
E_{L^2}&=&\sum_i\Bigg(\frac{\nu_i+1}{2}\log_2\frac{\nu_i+1}{2}-\frac{\nu_i-1}{2}\log_2\frac{\nu_i-1}{2}\Bigg),\nonumber\\
&&
\end{eqnarray}
where the symplectic eigenvalues $\nu_i$ are the square root of the
eigenvalues of $QP$ or $PQ$. It is important to note that all
symplectic eigenvalues must be greater than or equal to 1. We follow
the definition of the bipartite entanglement in reference
\cite{Plenio} in which they show the entanglement area law being
valid in a finite-ranged interaction harmonic system with arbitrary
dimension.

We now proceed to examine the relationship of the entropy $E_{L^2}$
and the boundary length $L$.  In fact, by using Fannes' inequality
\cite{Fannes}, we can argue that the bipartite entanglement entropy
of the many-body Hamiltonian $H_1$ and that of harmonic system $H_2$
are rather close for a finite block $L$ \cite{entbound}. Hence, the
harmonic system is a good model to study the area law of this exact
many-body system. In Fig. \ref{fig2}, the entropy $E_{L^2}$ against
the boundary length $L$ with different couplings $g$ are shown. The
entropy $E_{L^2}$ is directly proportional to the length $L$ in the
non-critical regime. Moreover, {\it in the case of the coupling $g$
closing to the critical coupling $g_{c}$ is also shown to satisfy
the area-law}. Therefore, the entropy of the bipartite entanglement
is still area-bounded in the vicinity of the QPT. For the
finite-size case, we consider the total number of sites are
$M{\times}M=6400$ in the system as shown in Fig. \ref{fig2}.  Our
numerical results, for this case are in a good agreement with the
infinite lattice case and also satisfy the area-law.

\begin{figure}[ht]
\includegraphics[height=6cm]{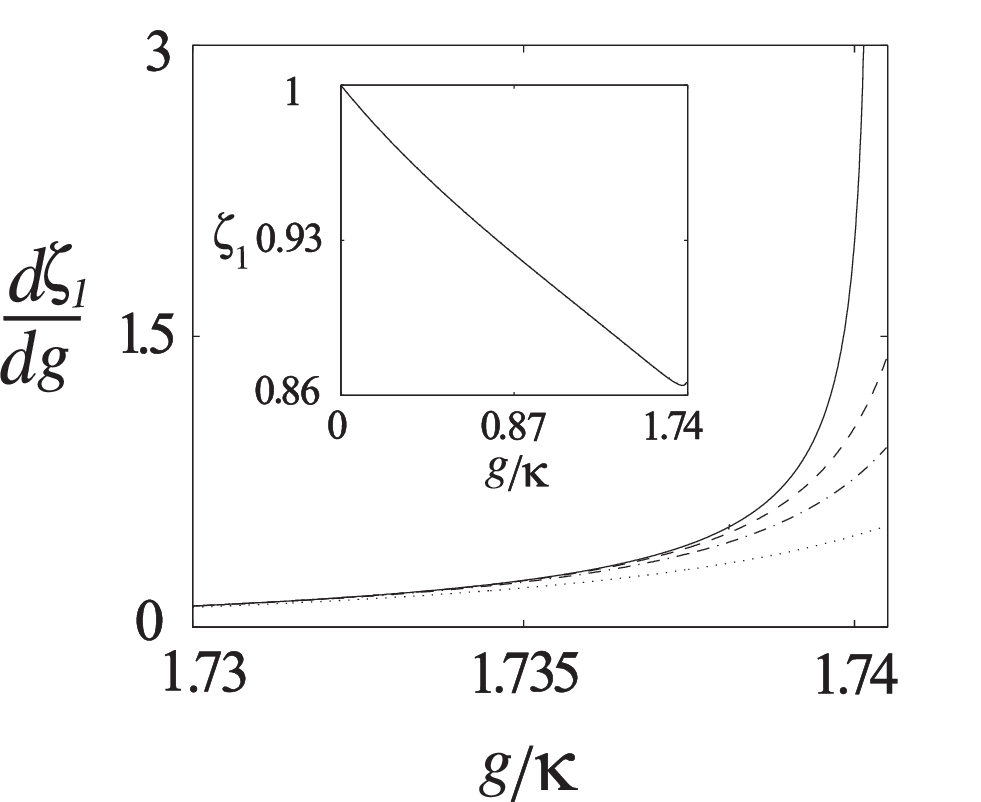}
\caption{ \label{fig3} The derivative $d\zeta_1/dg$ as a function of
$g/\kappa$ is shown with the same parameters to the previous figure.
The infinite case (solid line), $M=41$ (dashed line), 31 (dash-dot
line) and 21 (dotted line) are shown respectively. }
\end{figure}

\section{Two-site entanglement}
We shall now examine the entanglement between atoms trapped in  two
different sites. The problem of two-site entanglement reduces to
determining the entanglement between any two harmonic oscillators in
an ensemble of harmonic oscillators. Indeed, it is analogous to
determining two-mode entanglement in quantum optics \cite{Simon}.
The criterion of the inseparability of two-mode system \cite{Simon}
can be applied to the pure and the mixed states and hence it can be
used to evaluate two-site entanglement. The two-site reduced density
matrix $\rho^{(i,j)}$ can be constructed from the correlation
functions of these two sites $i$ and $j$ only. By applying several
local transformations, the reduced density matrix of a symmetric
Gaussian state can be written as \cite{Simon}
\begin{equation}
\label{symM} \rho^{(i,j)}= \left( {\begin{array}{clcr}
n & 0 &  c & 0 \\
0   &n&   0  & -c \\
c & 0 &  n & 0 \\
 0  &-c&   0  & n
\end{array}} \right),
\end{equation}
where $n=2(\langle{q^2_i}\rangle\langle{p^2_i}\rangle)^{1/2}
=2(\langle{q^2_j}\rangle\langle{p^2_j}\rangle)^{1/2}$ and
$c=2(-\langle{q_iq_j}\rangle\langle{p_ip_j}\rangle)^{1/2}$. This
two-site entanglement parameter can be defined as
$\zeta_{|i-j|}=n-c$ \cite{Ng}. If $\zeta_{|i-j|}$ is below one, then
it is said to be entangled. Moreover, this parameter can be used to
evaluate the amount of entanglement of formation (EOF)
\cite{Giedke}.  EOF is a function of the parameter $\zeta_{|i-j|}$
for $0<\zeta_{|i-j|}<1$ \cite{Giedke,Ng}.

In the infinite lattice limit, we can treat our system as a
symmetric Gaussian state due to the preservation of translational
symmetry. We are therefore able to evaluate the EOF through the
two-mode entanglement parameter in this case. In the finite-size
case, we investigate the oscillators locating at the center in the
square lattice and the total number of oscillators are odd. This
Gaussian state is nearly symmetric if the system size is large
enough. We study the two-site entanglement parameter $\zeta_1$ of
two adjacent sites. It decreases as the strength $g$ increases as
shown in the inset in Fig. \ref{fig3} but the minimum is not formed
at the critical point. This means that the two-site entanglement can
be controlled by the strength $g$. In addition, we found that it is
only the nearest neighbors that are entangled. The quantum
entanglement is thus very short-ranged indeed. In fact, this feature
coincides with the intuition of the area-dependent entropy. It is
because the oscillators in the ``inside'' region entangle with the
oscillators in the boundary. The entropy is thus at most
proportional to the boundary $L$.

In order to examine how the ground state changes close to a QPT
\cite{Osterloh}, we numerically examine the first derivative of the
two-site entanglement $d\zeta_1/dg$ of the nearest neighbor. In Fig.
\ref{fig3}, we see $d\zeta_1/dg$ diverging as the QPT is approached
in the infinite lattice.  In the case of a finite lattice there is
still an increase, albeit a finite one.

\section{Conclusion}
We have investigated the entanglement content in the spin waves of
the two-component condensates in a 2D square lattice with the
finite-ranged interaction close to a QPT. We have shown that the
entropy satisfies the area-law and examined the scaling of two-site
entanglement in this finite-ranged interaction system.  These
features of the system should be addressable by recently developed
experimental techniques that give access ti atom-atom correlation
function. Moreover, the physical realization of harmonic chain may
lead to applications in quantum information science \cite{Plenio2}.

H.T.N. is grateful to M. Plenio and J. Eisert for the discussion.
H.T.N. thanks the financial support of the Croucher Foundation and
K.B. thanks the Royal Society and Wolfson Foundation for support.


\begin{thebibliography}{99}
\bibitem{entbound}
By Fannes' inequality, the difference of two entropies are given by
\begin{eqnarray}
\label{entineq}
|E_{L^2}(\rho_1)-E_{L^2}(\rho_2)|{\leq}2D(\rho_1,\rho_2)L^2\log{N}\nonumber\\
-2D(\rho_1,\rho_2)\log{2D(\rho_1,\rho_2)},
\end{eqnarray}
where $D(\rho_1,\rho_2)$  is the trace distance between the density
matrices, $\rho_1$ and $\rho_2$ are the reduced density matrices of
the Hamiltonians $H_1$ and $H_2$ respectively.  We have truncated
the Hilbert space of $H_2$ to $N^{L^2}$ dimensions. For the
non-increasing property of the trace distance under partial trace
and relation of trace distance between the fidelity of two pure
states \cite{Nielsen}, we have
\begin{eqnarray}
D(\rho_1,\rho_2)&{\leq}&D(|\psi\rangle,|\phi\rangle),\\
&=&\sqrt{1-F^2(|\psi\rangle,|\phi\rangle)}, \label{Fidelityinq}
\end{eqnarray}
where $|\psi\rangle$ and $|\phi\rangle$ are the ground state of the
Hamiltonians $H_1$ and $H_2$ respectively. By the perturbation
theory, the leading approximation of this ground state of $H_1$ can
be found as
$|\psi\rangle{\approx}(|\phi\rangle+\epsilon|\psi^{(1)}\rangle)/\sqrt{1+\epsilon^2}$,
$\epsilon$ is small parameter of the $N^{-1}$ correction to the
ground state in the order of $\mathcal{O}(N^{-1})$ and
$|\psi^{(1)}\rangle$ is the first order of this ground state. The
fidelity $F(\rho_1,\rho_2)$ is about $1-\epsilon^2/2$, so it can be
shown that $D(\rho_1,\rho_2)$ is less than $\epsilon$ from
(\ref{Fidelityinq}). Therefore,  we can see that the  difference of
two entropies is bounded by a small value (\ref{entineq}) as long as
$N$ is sufficiently larger than $L^2\log{N}$.  The infinite-size
limit can be attained if we first take $N$ to be infinity and then
take $L$ to be infinity.
\end{thebibliography}
\end{document}